\documentclass[preprint,12pt]{elsarticle}

\usepackage{amssymb}
\usepackage{graphics}
\usepackage{graphicx}
\usepackage[edges]{forest}
\usepackage{array, colortbl}
\usepackage{forest}
\usepackage[utf8]{inputenc}
\usepackage{verbatim}
\usepackage{tabularx}
\usepackage{multirow}
\usepackage{supertabular}
\usepackage{longtable}
\setcitestyle{square,numbers,unsrt}
\usepackage{url}
\usepackage{cuted}
\usepackage{caption}
\usepackage{subcaption}
\usepackage{framed} 
\usepackage{multicol} 
\usepackage{ulem}
\usepackage{amsmath}
\usepackage{amsthm}
\usepackage{nomencl} 
\makenomenclature
\renewcommand*\nompreamble{\begin{multicols}{2}}
\renewcommand*\nompostamble{\end{multicols}}

\graphicspath{ {./Figures/} }
\def\BibTeX{{\rm B\kern-.05em{\sc i\kern-.025em b}\kern-.08em
    T\kern-.1667em\lower.7ex\hbox{E}\kern-.125emX}}

\newcolumntype{P}[1]{>{\centering\arraybackslash}p{#1}}
\newcolumntype{M}[1]{>{\centering\arraybackslash}m{#1}}
\newcolumntype{N}{@{}m{0pt}@{}}
  
\def\tsc#1{\csdef{#1}{\textsc{\lowercase{#1}}\xspace}}
\tsc{WGM}
\tsc{QE}

\journal{}

\begin{document}

\begin{frontmatter}

\title {A geographic information system-based modelling, analysing and visualising of low voltage networks: The potential of demand time-shifting in the power quality improvement}

\author{Tomislav Antić}
\cormark[*]
\fnmark[*]
\ead{tomislav.antic@fer.hr}

\author{Tomislav Capuder}
\ead{tomislav.capuder@fer.hr}

\affiliation{organization={University of Zagreb Faculty of Electrical Engineering and Computing, Department of Energy and Power Systems},
            city={Zagreb},
            country={Croatia}}

\cortext[1]{Corresponding author}

\begin{abstract}
The challenges of power quality are an emerging topic for the past couple of years due to massive changes occurring in low voltage distribution networks, being even more emphasized in the years marked by the novel COVID-19 disease affecting people's behaviour and energy crisis increasing the awareness and need of end-users energy independence. Both of these phenomena additionally stress the need for changes in the planning and operation of distribution networks as the traditional consumption patterns of the end-users are significantly different. To overcome these challenges it is necessary to develop tools and methods that will help Distribution System Operators (DSOs). In this paper, we present a geographic information system (GIS)-based tool that, by using open source technologies, identifies and removes errors both in the GIS data, representing a distribution network, and in the consumption data collected from the smart meters. After processing the initial data, a mathematical model of the network is created, and the impact of COVID-19-related scenarios on power quality (PQ) indicators voltage magnitude, voltage unbalance factor (VUF), and total voltage harmonic distortion (THD$_u$) are calculated using the developed harmonic analysis extension of the pandapower simulation tool. The analyses are run on a real-world low voltage network and real consumption data for different periods reflecting different COVID-19-related periods. The results of simulations are visualised using a GIS tool, and based on the results, time periods that are most affected by the change of end-users characteristial behaviour are detected. The potential of the end-users in the PQ improvement is investigated and an algorithm that shifts consumption to more adequate time periods is implemented. After modifying the consumption curve, power quality analysis is made for newly created scenarios. The results show that the pandemic negatively affect all analysed PQ indicators since the change in the average value of PQ disturbances increased both during the hard and post-lockdown period. The time-shifting of consumption shows significant potential in how the end-users can not only reduce their own energy costs but create power quality benefits by reducing all relevant indicators.
\end{abstract}

\begin{highlights}
\item Integrated tool for modelling, analysing, and visualising a network is presented
\item Simulations are run to determine the impact of end-user pattern consumption change on power quality (PQ)
\item The results show negative impact of passive end-user demand changes on PQ indicators
\item Demand time-shifting is shown to be an efficient method for reducing PQ disturbances
\end{highlights}


\end{frontmatter}


\section{Introduction}
\subsection{Impact of COVID-19 on power systems}
Besides the well-known technical challenges caused by the continuous increase of LC technologies share and their uncoordinated installation and operation \cite{TORRES2022117718,WANG2019113927}, in 2020, 2021, and part of 2022 the COVID-19 pandemic has created numerous changes in the planning and operation of power systems and new challenges for system operators. The analysis of the German and other European power systems showed that the pandemic increased the existing power system's uncertainty despite the kept high level of security. The authors in \cite{WANG2020115735} investigated the impact of COVID-19 on the CO$_2$ emissions and showed that in 2020 the emissions decreased in the industry, transportation, and construction sector. The authors in \cite{RUAN2021116354, COSTA2021107172} analyse the impact of the pandemic on electricity markets and market operations and show that both electricity prices and power demand were seriously affected by the reaction to the virus. Results of a detailed analysis of the impact of COVID-19 on the behaviour of end-users show that electricity consumption of residential low voltage (LV) end-users has significantly increased \cite{MAHFUZALAM2021106832, ROULEAU2021116565}. The focus of already published research on the COVID-19 impact on LV distribution networks is mostly on the change in consumption and its correlation with environmental, technical, and market aspects of planning and operation of distribution networks. However, the existing literature does not address the impact of the pandemic on power quality and the changes in people's behaviour in the post-pandemic period and if their potential in providing flexibility services changes. To overcome the identified gap, we present a detailed analysis of power quality in the LV distribution network over different pandemic-related periods and observe if it is possible to improve deterioration only by shifting demand with the minimum disruption of end-users' comfort.

\subsection{Open-source power system simulations}
The use of power system simulation tools developed using different programming languages is continuously increasing. The development of pandapwoer, the Python-based open-source tool for different analyses of steady-state power systems is presented in \cite{8344496}. The mathematical model and the pandapower-based tool for harmonic analysis in both balanced and unbalanced radial distribution networks were presented in \cite{ANTIC2022107935}. Another open-source platform for modelling, control, and simulations of local energy systems, OPEN, was described in \cite{MORSTYN2020115397}. MATPOWER, an open-source Matlab-based power system simulation package that provides a high-level set of power flow, optimal power flow (OPF), and other tools targeted toward researchers, educators, and students are described in detail by the authors in \cite{5491276}. Major benefits of open source tools include the possibility to modify and extend their functionalities and to integrate them with other tools and external databases. Such integration with smart meter measurements and geographic information system data is described in detail in the rest of the paper.

\subsection{The power of geographic information systems}
Due to the increasing complexity of distribution networks, it becomes insufficient to rely only on power system simulation tools. Such tools must be integrated with other tools, systems, and databases. One of the systems being more used in the planning and operation of smart distribution networks is the geographic information system (GIS), a system that contains both technical and geospatial attributes of distribution networks. The authors in \cite{GisOpenDSS} present an integration of QGIS, an open source GIS tool, with OpenDSS, an open source tool for power system simulations. Since GIS is characterized by georeferenced data, other applications of GIS-based tools include determining the PV hosting capacity or finding an optimal location for installing a PV power plant \cite{ZAMBRANOASANZA2021110853,8270364}, developing an optimal design of renewable-powered EV charging stations \cite{HUANG2019113855}, or visualising the results of power flow calculations \cite{9248930}. A potential disadvantage of using GIS data in its integration with the power system analysis tool is the high probability of different errors in the initial data set. Those errors often make the data unusable in its initial form and further processing and editing are required \cite{8493353}. Other than GIS data, data collected from smart meters is often considered the key in different distribution network analyses \cite{KU2022118539, QI2020115708} or forecasts \cite{FEKRI2021116177}. A detailed analysis of the potential use and advantages of smart meters in the planning and operation of smart networks is presented in \cite{YILDIZ2017402,8322199}. Unlike the other papers that are mostly focused on the single use of GIS in the planning and operation of power systems, in this paper, GIS is applied in editing data representing LV network elements and visualisation of the results of power system simulations. On top of that, we present a detailed step-by-step guide for detecting and editing specific geospatial errors that prevent the direct use of GIS data.

\subsection{Mitigation of power quality issues}
To prevent power quality disturbances and other technical problems in distribution networks, it is becoming more important to investigate approaches for mitigating these issues. Solutions can be oriented on the utilization of physical devices such as inverters that help in the mitigation of harmonic distortion \cite{6408380, 8637800} or re-phasing switches that mitigate network unbalances \cite{CHAMINDABANDARA2020116022}, the potential of end-users where they can help in the improvement of power quality \cite{THOMAS2020114314} and in avoiding harmonics \cite{CICEK2021106528} under multiple demand response schemes created based on the profiles created in the laboratory environment, or on their combination as described in \cite{VIJAYAN2021117361}. Besides observing only the potential of end-users' increase or decrease in electricity demand, the improvement of networks' technical conditions can be achieved by the time-shifting of demand. However, this approach is in most cases applied only on the improvement of one value, e.g., voltage magnitude along the feeder \cite{PAPAIOANNOU2013540}. As seen in the literature review, papers often focus on the improvement of the limited set of technical quantities and solutions are often tested on benchmark networks or in a controlled environment. We propose a demand time-shifting algorithm aimed at the comprehensive improvement of PQ indicators. The mathematical model of the algorithm is tested on real-world data - demand measurements collected from smart meters and real-world LV network topology information.

\subsection{Contributions}
In this paper, we propose an integrated open source tool used for processing and editing GIS and smart meter data, power quality analyses, and visualising the results of the simulations. QGIS, Python, and SQL are used for processing and editing GIS data, smart meter data is processed using Python and SQL, power quality is analysed using the pandapower tool and its developed harmonic calculations extension, while the results are visualised using QGIS. Additionally, Python is used for the detection of time periods that were most affected by the pandemic-related changes. After the detection, a demand time-shifting algorithm is implemented in order to investigate if it is possible to reduce the PQ deterioration only by using end-users potential and without the investment in the additional physical devices. Power quality analyses in this paper are limited only to voltage magnitude, VUF, and THD since these are the only disturbances that can be calculated using the presented model. Other disturbances cannot be detected by quasi-static time series simulations using pandapower or any other tool with the same set of functionalities and their analysis is out of the scope.

To summarise, based on the literature review and identified research gaps, the proposed contributions of this paper are:
\begin{itemize}
    \item An integrated open source tool used for editing and processing GIS and smart meters data, creating LV network models, performing power quality analyses, and the visualisation of the simulations results.
    \item A comprehensive analysis of PQ indicators in the real-world LV network during periods related to realistic end-users consumption pattern changes, i.e. COVID-19-related pre-pandemic, pandemic, and post-pandemic periods. Calculated values are compared to limitations in European standards and the impact of the pandemic on the distortion of power quality is investigated.
    \item Implementation of the end-users demand time-shifting scheme based on the detection of PQ distorted intervals. The developed methodology reduces values of PQ indicators and in some cases even improves PQ and, unlike the papers that rely only on the possibility of end-users to increase or decrease their consumption when needed, results in the effect of less discomforting and more acceptable time-shifting of end-users' consumption on the PQ improvement.

\end{itemize}

The rest of the paper is organized as follows: Section \ref{sec:data_edit}  gives an explanation of detected errors in GIS and smart meters data and methods used for removing the errors, while the results of analyses defined for a real-world Croatian case study are defined in Section \ref{sec:case_studies}. Section \ref{sec:improvement} presents the methodology and the results after the implementation of an algorithm used for the improvement of PQ. Finally, conclusion is presented in Section \ref{sec:conclusion}.

\section{Tool description}\label{sec:data_edit}
Using both GIS and smart meters data presents great potential in the improvement of the planning and operation of smart distribution networks and the increase of the network's observability. The problem with this data is that in a high number of cases it cannot be used in its initial form, which presents an obstacle for Distribution System Operators (DSOs). Also, not all data about distribution networks is stored in one system or database. Therefore it is necessary to develop an integrated tool that enables the communication between several systems and databases, that has the possibility of removing errors and preparing data for further analyses.

In the first step, the open source-based tool presented in this paper is used for editing the data and removing the errors in the initial data set. Features of open source technologies QGIS, Python programming language, and PostgreSQL with the postgis extension, are applied for processing the initial GIS and smart meters data. GIS data contains technical and geospatial information about the Croatian real-world LV network, while smart meters data contains information about the end-users' consumption. The focus in this paper is put on data representing a LV network since the main focus was to investigate the potential of residential end-users in the improvement of technical conditions in distribution networks. In case of observing MV networks, the presented tool can be used for processing the GIS data and in case of identifying new errors it can be updated in order to successfully solve all detected problems.

\subsection{GIS data}
In the process of identifying the errors in the initial GIS data set, several errors were detected, and afterward, the data was edited in order to enable using it in the creation of the network's mathematical model:
\begin{itemize}
    \item Continuity of a polyline
    \item Unknown beginning and ending node of an LV cable/line
    \item Disconnection of an LV cable/line and an MV/LV substation
    \item Disconnection of an LV cable/line and an LV switch cabinet
    \item Unknown technical attributes of an LV cable/line
    \item LV cable/line without known ending node
    \item Redundance of point objects 
\end{itemize}

\begin{figure}[htbp]
    \centering
    \includegraphics[width=\columnwidth]{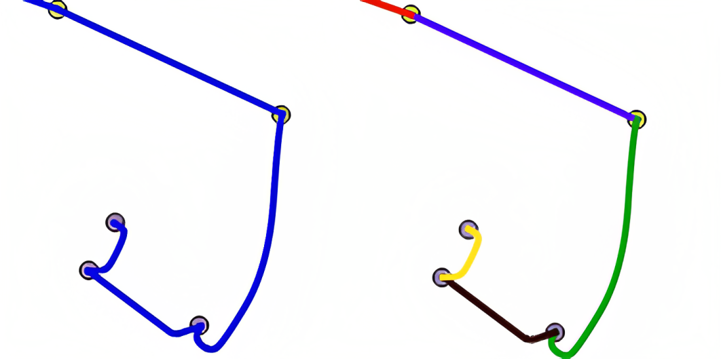}
    \caption{Dividing a line object into segments}
    \label{fig:before_after_line}
\end{figure}

Each polyline that represent the main line of the observed distribution network feeds additional branches of a distribution network. Main lines and branches are connected at a point object called connection point. Even though main line is topologically connected with branches, without breaking a polyline into smaller parts, i.e., segments, determining the beginning and the ending node of each line will not be possible. Figure \ref{fig:before_after_line} shows an example of dividing a line object into segments, where each colour represents an individual segment created by using the developed tool.

Some point objects, e.g., connection points or points that represent end-users, are located at the beginning or the ending node of a line object, i.e., some point objects are topologically connected with line objects representing an LV line/cable. In those cases, based on the equality of coordinates, the beginning and ending nodes of each line object are determined. This change cannot be visualised since only attributes of a line object are changed in a database so they could be used in the creation of a mathematical model of a network based on the edited GIS data.

\begin{figure}[htbp]
    \centering
    \includegraphics[width=\columnwidth]{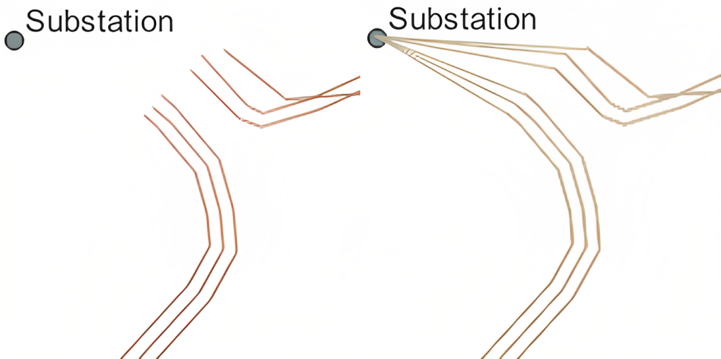}
    \caption{Connection of a point with line objects}
    \label{fig:before_after_sub}
\end{figure}

Some point objects are not located at the beginning or the end of a line, but still, need to be connected with lines. An example of those objects are points representing MV/LV substations or LV switch cabinets. Depending on which node, the beginning or the ending, of a line is unknown, the distance between the first or last coordinate of a line and a point object representing a substation or an electric switch cabinet is determined, and if it is within the radius defined by a user, the point object is attached as a beginning or ending node of a line. Figure \ref{fig:before_after_sub} shows an example of a topological connection of a point object representing a substation with line objects representing lines and cables.

\begin{figure}[htbp]
    \centering
    \includegraphics[width=\columnwidth]{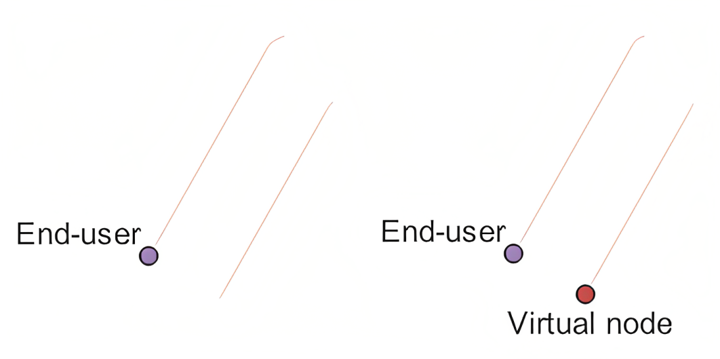}
    \caption{Creating a point object}
    \label{fig:virt_node_before_after}
\end{figure}

After all point objects are processed there are still some lines without attached beginning or/and ending nodes. In those cases the missing node is replaced with a point object called a virtual node, which is an element without a connected end-user, i.e., the value of the load at that node is equal to zero. An example of creating a point object representing the virtual node element is shown in Figure \ref{fig:virt_node_before_after}.

\begin{figure}[htbp]
    \centering
    \includegraphics[width=\columnwidth]{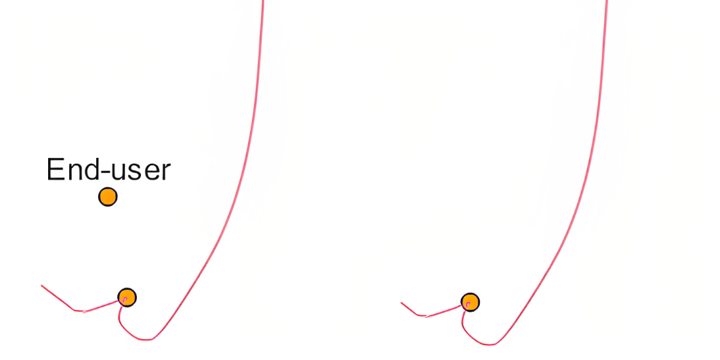}
    \caption{Deletion of topologically unconnected node}
    \label{fig:extra_node}
\end{figure}

Some point objects cannot be connected with LV cables/lines since they do not belong to the observed network, i.e., these point objects are not a part of the network and they were input by an accident. Those objects must be deleted from the database and the GIS data set, since they will cause errors in further calculations, e.g., an unbalanced power flow calculation becomes impossible. Figure \ref{fig:extra_node} shows an example of deletion of an extra node, i.e., of a node representing an end-user that is not connected with the rest of a network.

In the last step of GIS data editing, unknown technical attributes of LV lines/cables are replaced with values for the standard type of a line or a cable that is most common in a database. 

\subsection{Smart meters data}
Similar to the error detection in the initial GIS data set, errors are identified in the collected smart meters data:
\begin{itemize}
    \item outlier values
    \item zero values
    \item missing values
\end{itemize}

Outlier values are defined as the values that significantly differ from the other consumption data, e.g., the consumption of an end-user at one time period is 1000 times larger than in the previous one. In some time periods, the value of consumption is equal to zero. Even though it is possible that consumption lowers at certain periods, there are still several devices in each household that consume energy for basic functions. In some periods, due to a loss of connection between the smart meter and a database or any other obstacle, values of consumption are not stored in the database. Those values are defined as missing ones. All detected errors are resolved in a way that an outlier, zero, or a missing value in a certain period is replaced with the value of consumption in a previous time period. If there are multiple consecutive time periods with some of the detected errors, there is a significant number of intervals in which the initial data cannot be used, or detected errors could not be corrected, the whole consumption data collected from a certain smart meter is neglected. For the purpose of the analysis in this paper, such data is replaced with consumption collected from other smart meters, modified with a randomly determined factor, since it is not expected that two end-users have identical consumption in each time period. An example of the smart meter data editing is shown in Table \ref{tab:smart_meter_edit}.

\begin{table}[]
\caption{Example of the smart meter data editing}
\centering
\resizebox{\columnwidth}{!}{
\begin{tabular}{c|ccclccc|}
\cline{2-8}
                            & \multicolumn{7}{c|}{Power (W)}                                                                                                                                                                                                                                                                                                                                                                                                                                                                                                    \\ \cline{2-8} 
                            & \multicolumn{1}{c|}{\begin{tabular}[c]{@{}c@{}}Outlier\\ value\end{tabular}} & \multicolumn{1}{c|}{\begin{tabular}[c]{@{}c@{}}Zero \\ value\end{tabular}} & \multicolumn{1}{c|}{\begin{tabular}[c]{@{}c@{}}Missing\\ value\end{tabular}} & \multicolumn{1}{l|}{\multirow{6}{*}{$\Rightarrow$}} & \multicolumn{1}{c|}{\begin{tabular}[c]{@{}c@{}}Outlier\\ value\end{tabular}} & \multicolumn{1}{c|}{\begin{tabular}[c]{@{}c@{}}Zero \\ value\end{tabular}} & \begin{tabular}[c]{@{}c@{}}Missing\\ value\end{tabular} \\ \cline{1-4} \cline{6-8} 
\multicolumn{1}{|c|}{Time}  & \multicolumn{3}{c|}{Initial data}                                                                                                                                                                                                        & \multicolumn{1}{l|}{}                                            & \multicolumn{3}{c|}{Edited data}                                                                                                                                                                                    \\ \cline{1-4} \cline{6-8} 
\multicolumn{1}{|c|}{00:00} & \multicolumn{1}{c|}{575}                                                     & \multicolumn{1}{c|}{575}                                                   & \multicolumn{1}{c|}{575}                                                     & \multicolumn{1}{l|}{}                                            & \multicolumn{1}{c|}{575}                                                     & \multicolumn{1}{c|}{575}                                                   & 575                                                     \\ \cline{1-4} \cline{6-8} 
\multicolumn{1}{|c|}{00:15} & \multicolumn{1}{c|}{583}                                                     & \multicolumn{1}{c|}{583}                                                   & \multicolumn{1}{c|}{583}                                                     & \multicolumn{1}{l|}{}                                            & \multicolumn{1}{c|}{583}                                                     & \multicolumn{1}{c|}{583}                                                   & 583                                                     \\ \cline{1-4} \cline{6-8} 
\multicolumn{1}{|c|}{00:30} & \multicolumn{1}{c|}{50000}                                                   & \multicolumn{1}{c|}{0}                                                     & \multicolumn{1}{c|}{-}                                                       & \multicolumn{1}{l|}{}                                            & \multicolumn{1}{c|}{583}                                                     & \multicolumn{1}{c|}{583}                                                   & 583                                                     \\ \cline{1-4} \cline{6-8} 
\multicolumn{1}{|c|}{00:45} & \multicolumn{1}{c|}{580}                                                     & \multicolumn{1}{c|}{580}                                                   & \multicolumn{1}{c|}{580}                                                     & \multicolumn{1}{l|}{}                                            & \multicolumn{1}{c|}{580}                                                     & \multicolumn{1}{c|}{580}                                                   & 580                                                     \\ \cline{1-4} \cline{6-8} 
\end{tabular}}
\label{tab:smart_meter_edit}
\end{table}

\subsection{Power quality analysis and visualisation of the results}
After all the data is processed and edited and all errors are removed from the initial data sets, prior to calculations it is necessary to merge the data related to the end-users consumption with the GIS data. Since end-users consumption and GIS data are stored in two unrelated databases, the communication between the two systems is established as a part of the developed tool. After the merging of the data, all prerequisites for analyses of LV distribution networks are satisfied.

The second step of the developed tool includes the creation of a mathematical model of an LV network using pandapower, an open source Python-based library for power systems calculation \cite{8344496}. Additionally, one of the latest pandapower extensions presented in \cite{ANTIC2022107935} is used for the unbalanced harmonic analysis. LV network elements such as nodes, lines, transformers, and loads (determined by locations of end-users) are created from the edited GIS data, while the value of each load's active power is defined from the smart meter data. Since only active consumption is measured, reactive power of loads is calculated with power factor, randomly chosen from the $[0.95-1]$ interval. Once the network with all its elements is created, pandapower is used for time-series simulations and calculation of voltage magnitude, voltage unbalance factor (VUF), and total voltage harmonic distortion (THD$_u$). Consumption measurements are collected in the 15-minute interval over one week and they represent the average value of consumption for every end-user in every 15-minute interval. Because of that, calculated values of PQ indicators voltage magnitude, VUF, and THD are also average 15-minute values and not values captured at a certain moment. Therefore, the focus of analyses is on the quasi-static time series three-phase harmonic power flow and PQ indicators that are meaningful in the defined time interval. Other indicators such as transients require the installation of additional monitoring devices and cannot be calculated with the used approach and are out of the scope of the paper.

Once the simulations are completed, the results are stored in an external database and used for visualisation using QGIS. In this paper, visualisation is intended to show the change of PQ indicators during the pandemic with respect to the pre-pandemic period, i.e., we investigate and visualise how much did PQ deteriorate or improve due to the changes caused by differences in end-users behaviour during different pandemic-defined scenarios. Results of the scenario that considers the implementation of the consumption time-shifting are also presented, which enables a simple and intuitive assessment of the end-users potential in the general improvement of power quality disturbances.

\begin{figure*}[htbp]
    \centering
    \includegraphics[width=\textwidth]{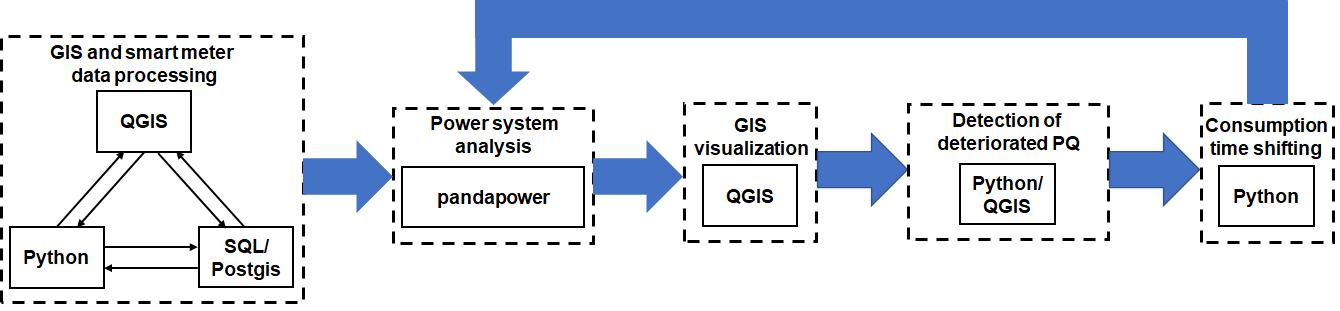}
    \caption{Integrated tool description}
    \label{fig:toolflow}
\end{figure*}

Tools used in each part of the analyses and the visualisation in this paper, and the developed integrated tool as a whole are presented in Figure \ref{fig:toolflow}.

\section{Case Study and Results}\label{sec:case_studies}
\subsection{Case Study}
\begin{figure}[htbp]
    \centering
    \includegraphics[width =\columnwidth]{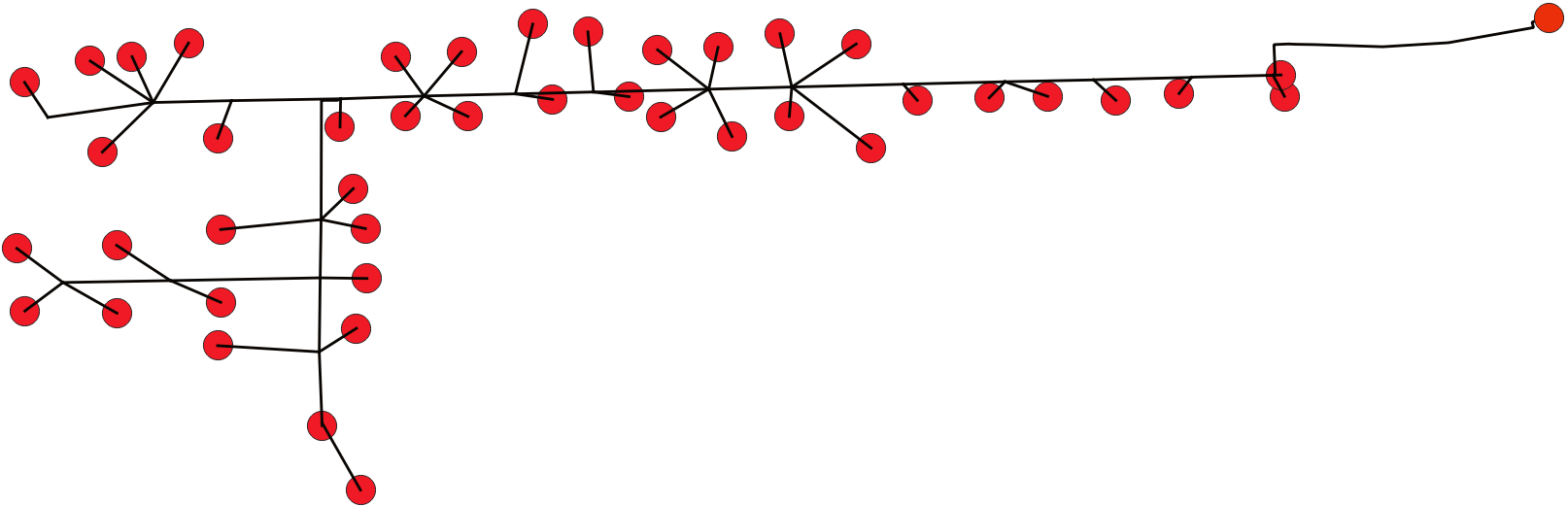}
    \caption{Modelled LV feeder}
    \label{fig:feeder}
\end{figure}

An LV network used for the analysis of the impact of different periods related to COVID-19 on the power system is a part of the real-world Croatian distribution network. An urban network consists of an MV/LV transformer, 10 feeders, 179 nodes, from which 139 represent end-users, and 177 overhead lines and underground cables. However, due to the visibility and clarity of the representation of the results, only one LV feeder, with 66 nodes, 43 three-phase and single-phase connected users, and 64 lines, is modelled. The modelled LV feeder is presented in Figure \ref{fig:feeder}. 

Three weeks that represent pre-lockdown, hard lockdown, and post-lockdown periods are chosen to minimise the season's impact on electricity consumption. The first period is set in late winter when the temperatures are not so low, and end-users heat less than in December or January. The second period is set in spring and is characterized by conditions that do not require heating or cooling. Finally, the third period is set in late summer, when the temperatures do not go as high as in July and August, and therefore end-users do not use air conditioners as often as in other summer months. Therefore, consumption and consequentially PQ indicators should not differ so much in the chosen representative weeks. Even more important, consumption during each representative week has captured specific patterns and end-users' behaviour that characterized different periods during the pandemic. Moreover, observation of these specific weeks allows the analysis of the impact of the pandemic on technical conditions in a residential LV network and also presents the opportunity to investigate if some periods were more favourable in terms of the end-users' flexibility provision.

The data from representing weeks presents the cumulative active consumption of end-users connected to an urban LV network, which is converted to the more suitable active power data. The active consumption data is collected in 15-minute intervals. Since the smart meters installed in the network defined in the case study do not collect the reactive consumption data, reactive power is calculated from active power and power factor whose value is chosen with uniform randomization from the $[0.95-1.0]$ interval. Together with the data defining the elements of an LV network, this data is used as input into pandapower and its harmonic calculation extension. 

\subsection{Results}\label{sec:results}
The results of simulations are visualised using the QGIS open source tool. Results show the average change of a PQ indicator during hard (Figure \ref{fig:PQ_hard}) and post-lockdown (Figure \ref{fig:PQ_post}) periods, compared to the pre-lockdown period. The change is calculated with eq. (\ref{eq:change}) where $PQ_{indicator}$ represents the value of voltage magnitude, VUF, or THD$_u$ during each observed COVID-19-related period. Finally, the average change for each PQ indicator in every node is calculated as a mean value of calculated $\Delta_{indicator}$ values in all periods. 

\begin{equation}
    \Delta_{indicator} = \frac{PQ_{indicator, hard/post} - PQ_{indicator, pre}}{PQ_{indicator, pre}}
    \label{eq:change}
\end{equation}

The defined method of calculating the change considers the pre-lockdown as a referent period that observed end-users' habits before the pandemic. Despite the opening of several business sectors and the relaxation of measures in most of countries during the post-lockdown period, the habits of end-users were changed, e.g., people are working from home more than in the pre-lockdown period. Therefore, the post-lockdown period is also characterized by increased electricity consumption in LV networks. Hence, the change in the value of each PQ indicator is observed with respect to the referent pre-lockdown period. It is important to mention that the change of the value of a certain PQ indicator does not mean that it increased or decreased for the calculated amount, e.g., if the value of THD$_u$ during the hard lockdown period increased to 3\% from 2\% in the pre-lockdown period, the increase of 50\% and not 1\% will be visualised using the developed tool. The reason why we focused only on these three PQ indicators in our analyses is the interval in which their values can be calculated. While there are other indicators such as transients, voltage swells or sags, these disturbances are characterized by a short duration or a milliseconds time frame in which they can be detected. There are two main reasons why it is not possible to consider them in analyses - the first being the need for additional measuring equipment since the only available data are the consumption measurements. The second reason is the limitation in terms of possible analyses conducted by the used power system simulation tool. However, other tools used in quasi-static power system simulations share this limitation and make the analyses of dynamic disturbances impossible and not realistic.

\subsubsection{Pre-lockdown period}

To better evaluate the change of values of PQ indicators during hard and post-lockdown periods, voltage magnitude, VUF, and THD$_u$ are also analysed for the pre-lockdown period in order to calculate values of PQ indicators in the base scenario and determine if there are any violations of defined PQ limitations or if values of observed PQ indicators come close to the threshold values. The minimum value of voltage magnitude is 0.933 p.u., while the maximum is 1.001 p.u. Since most of the European standards and national grid codes define an allowed interval of [0.9-1.1 p.u.] for the value of voltage magnitude, calculated values are not outside the allowed interval in any period during the observed week. However, the minimum value during the pre-lockdown period comes close to the lower bound of the allowed voltage magnitude interval, meaning that any further increase in consumption caused by the pandemic or electrification of heating and transport could cause a violation of defined limitations. The largest VUF value during the pre-lockdown period is equal to 0.539\%, which is below the 2\% threshold (3\% in networks with predominant single-phase connected loads). Even though most of the end-users' devices are single-phase connected to a network none of the devices have power that would create unbalance problems in a network. Single-phase connected devices lead to increased unbalance, but the value of VUF still does not violate the defined limitations. However, installation of any load with large nominal power or a single-phase connected PV will additionally increase the difference between phase consumption and consequentially phase voltages, i.e., voltage unbalance. A similar situation is with THD$_u$. Even without the integration of LC units that are power electronics interfaced to a network, end-users in LV networks possess a high number of non-linear loads that increase harmonic pollution. The largest value of THD$_u$ in the observed network is 0.886\% which is significantly lower than the maximum allowed value which is equal to 8\%. Same as in the case of voltage magnitude and VUF, the increase in consumption and further integration of LC units will only contribute to the harmonic distortion in a network.

\subsubsection{Hard lockdown period}

\begin{figure}
     \centering
     \begin{subfigure}[b]{\columnwidth}
         \centering
         \includegraphics[width=\columnwidth]{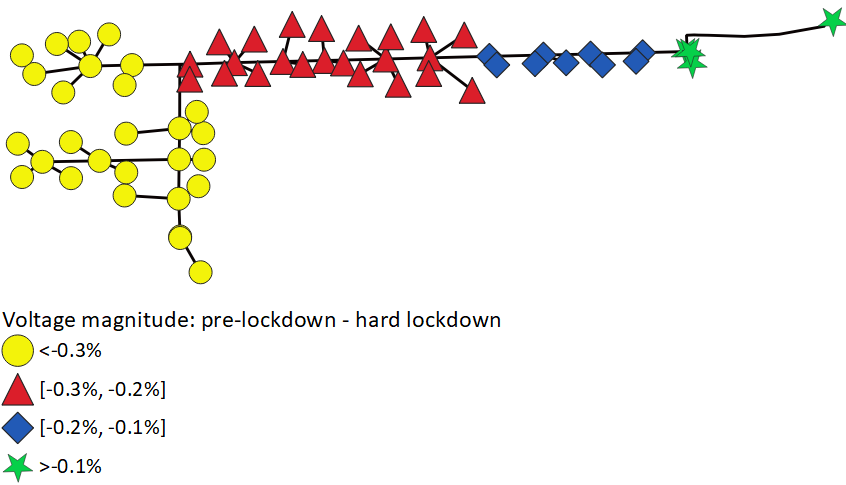}
         \caption{Change of voltage magnitude during hard lockdown}
         \label{fig:voltage_pre_hard}
     \end{subfigure}
     \hfill
     \begin{subfigure}[b]{\columnwidth}
         \centering
         \includegraphics[width=\columnwidth]{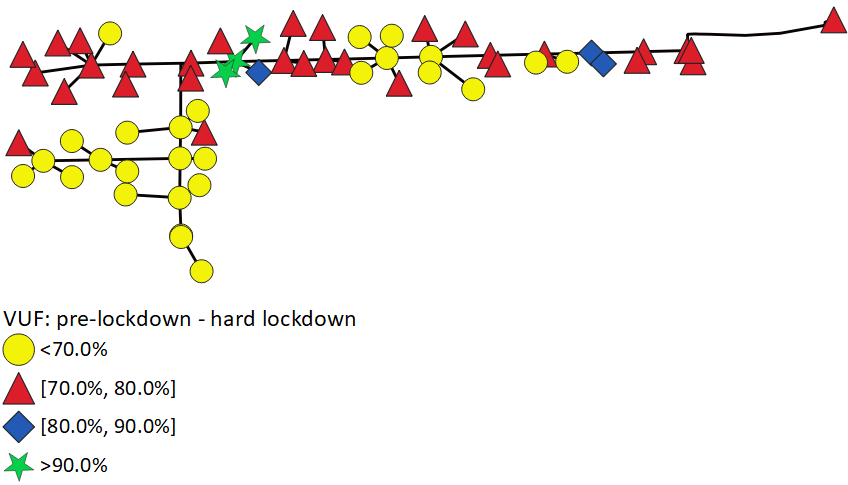}
         \caption{Change of VUF during hard lockdown}
         \label{fig:vuf_pre_hard}
     \end{subfigure}
     \hfill
     \begin{subfigure}[b]{\columnwidth}
         \centering
         \includegraphics[width=\columnwidth]{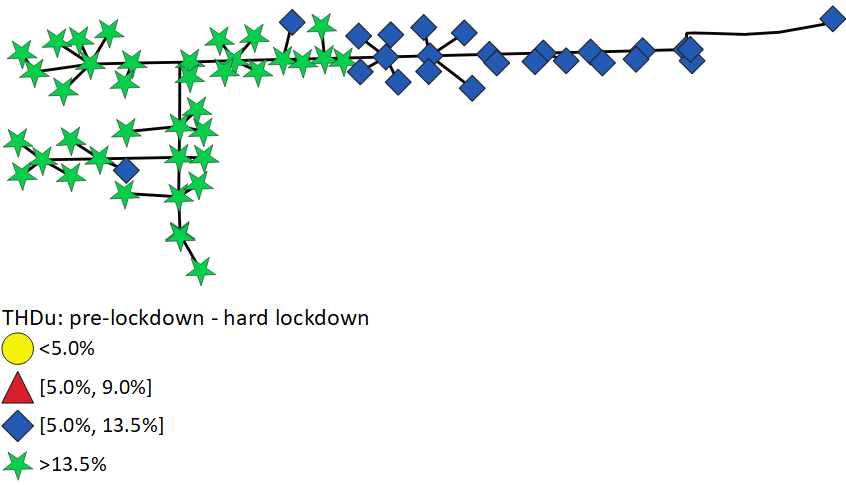}
         \caption{Change of THD during hard lockdown}
         \label{fig:thd_pre_hard}
     \end{subfigure}
        \caption{Change of PQ indicators during hard lockdown}
        \label{fig:PQ_hard}
\end{figure}

Figure \ref{fig:voltage_pre_hard} presents the average change of voltage magnitude in each node during the hard lockdown period compared to the referent, pre-lockdown period. The results show that the average value of voltage magnitude decreased in all nodes. The decrease is larger than 0.3\% at the nodes further from the referent node representing a substation. Voltage magnitude at first several nodes decreased for less than 0.1\%, while at all other nodes the decrease is in the range between 0.1\% and 0.3\%. Even though the value of voltage magnitude is close to the lowest value in the pre-lockdown period, change during the hard-lockdown period does not significantly decrease voltage magnitude. The lowest value is equal to 0.928 p.u., which is negligibly lower compared to the pre-lockdown period. Even though the results show that values of voltage magnitude do not violate the limitations even in the case of enlarged consumption, the decrease presents a concerning trend for DSOs. Since end-users' habits changed during the pandemic, higher consumption is expected in future scenarios and it must be considered in the planning and operation of distribution networks. Also, further and more rapid integration of LC units including heat pumps and EVs will continue the decrease of voltage magnitude and if not prevented, violations of limitations will almost certainly happen.

The change in average VUF value during the hard lockdown period, compared to the pre-lockdown periods is shown in Figure \ref{fig:vuf_pre_hard}. Unlike the average change of voltage magnitude, which was in the range of -0.327\% and -0.004\%, the change of VUF is significantly larger and is between 64.62\% and 106.39\%. In most nodes, the average change is lower than 70\% or between 70\% and 80\%, while in several nodes it is larger than 80\%. Even though the average change in each node seems very high and concerning, the maximum value of VUF in the hard lockdown period is equal to 0.659\%. The results of the analysis show that voltage unbalance increased due to enlarged consumption but it still did not lead to the violation of the threshold value. However, the distribution of consumption among phases is still close to symmetrical, which means that voltage unbalance still does not present a problem for DSOs. Same as with voltage magnitude, an increase of VUF presents a concerning trend which will in the future be more emphasized with the integration of additional single-phase consumption and production units.

Figure \ref{fig:thd_pre_hard} shows the impact of increased consumption during the hard lockdown period on harmonic distortion. Even though it is not the only factor in the calculation of THD$_u$, increased consumption causes higher current at the fundamental frequency and also higher current at higher frequencies since it is defined as the percentage of fundamental frequency current. Such increased current directly increases higher-order harmonic voltages, which together with already shown decreased voltage magnitude at fundamental frequency leads to higher values of THD$_u$ in the observed LV network. Compared to the pre-lockdown period, voltage distortion increased between 12.7\% and 14.22\% in all nodes, and in more than half of the nodes, the change is larger than 13.5\%. However, the largest value of THD$_u$ is 1.027\%, which is significantly lower than the 8\% threshold value. Despite the maximum value being lower than the limitation, any increase in harmonic distortion caused by higher consumption could present a potential problem for DSOs, since it is expected for consumption to continue to increase in the future, together with the share of power electronics.

\subsubsection{Post-lockdown period}

In the post-lockdown period analysed in this paper, most of the government's restrictions were abandoned, however, a significant number of end-users still remained working from home which continued the trend of increased consumption and changes related to PQ in residential LV networks.

\begin{figure}
     \centering
     \begin{subfigure}[b]{\columnwidth}
         \centering
         \includegraphics[width=\columnwidth]{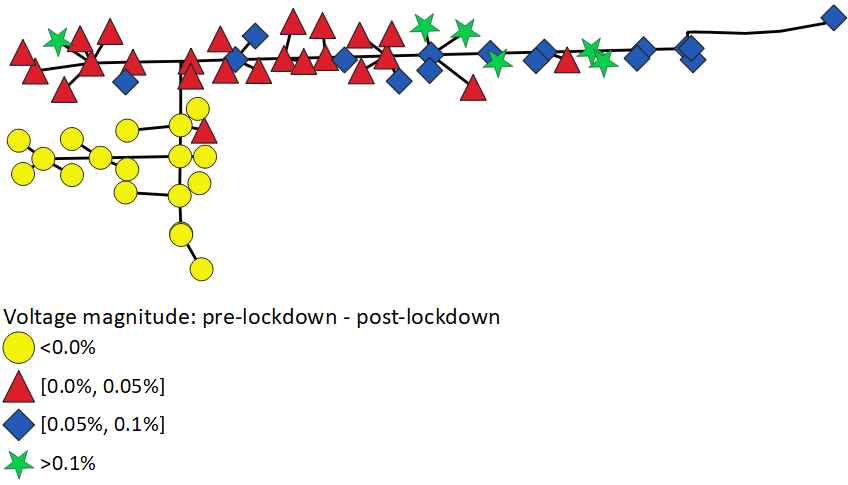}
         \caption{Change of voltage magnitude during post lockdown}
         \label{fig:voltage_pre_post}
     \end{subfigure}
     \hfill
     \begin{subfigure}[b]{\columnwidth}
         \centering
         \includegraphics[width=\columnwidth]{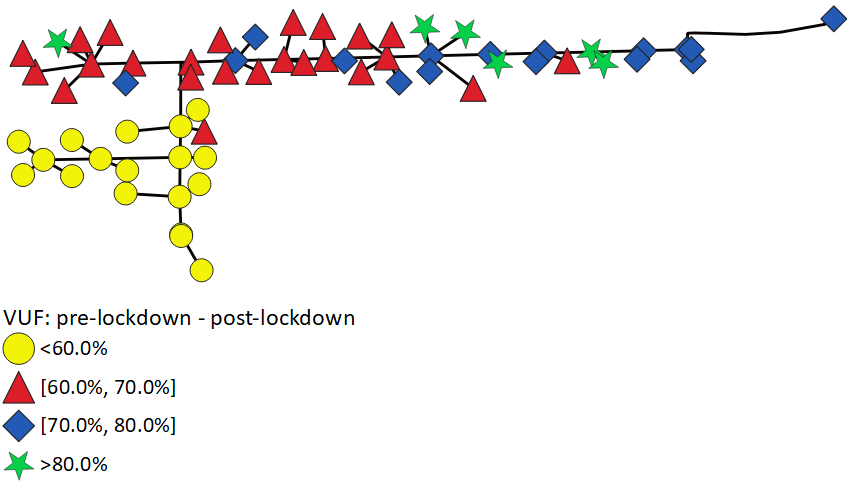}
         \caption{Change of VUF during post lockdown}
         \label{fig:vuf_pre_post}
     \end{subfigure}
     \hfill
     \begin{subfigure}[b]{\columnwidth}
         \centering
         \includegraphics[width=\columnwidth]{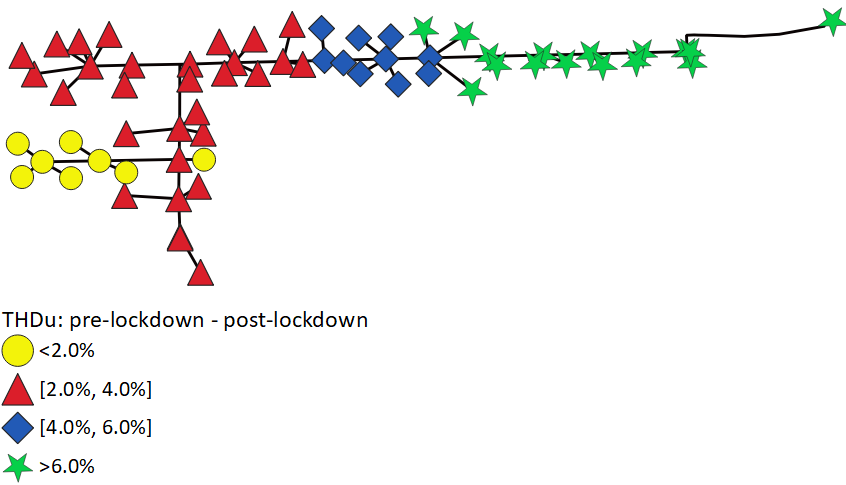}
         \caption{Change of THD$_u$ during post lockdown}
         \label{fig:thd_pre_post}
     \end{subfigure}
        \caption{Change of PQ indicators during post lockdown}
        \label{fig:PQ_post}
\end{figure}

Figure \ref{fig:voltage_pre_post} shows the change of voltage magnitude during the post-lockdown period, with respect to pre-lockdown. Compared to the hard-lockdown period, the average change of voltage magnitude in this scenario is lower and it even increased in some nodes, with the largest average increase of 0.124\%. The largest average decrease of voltage magnitude is -0.048\%, meaning that the average value of voltage magnitude is similar to the pre-lockdown period. Also, the minimum voltage magnitude is 0.939 p.u., which is negligibly larger than in the pre-lockdown period. Despite the fact that the behaviour of end-users changed, leading to more often working from home, the post-lockdown period is located in the late summer, almost fall, when consumption is traditionally smaller compared to winter periods, in which the pre-lockdown period is set. The pre-lockdown period is set in late winter when heating is still necessary, especially in continental, more cold parts, where the observed LV network is located. Therefore, the impact of seasonality is more emphasised than the post-pandemic-related changes in end-users' behaviour and the results are more similar to the pre-lockdown than the hard lockdown period.

The change of VUF during post-lockdown is shown in Figure \ref{fig:vuf_pre_post}. As mentioned before, the connection of large single-phase loads or single-phase connected generation units is more influential on voltage unbalance than seasonality or a scenario related to the pandemic. However, similar to during the hard lockdown period, the results show that there is a higher difference between phase consumption in the post-lockdown period. Compared to the pre-lockdown period, VUF increased in all nodes, but the change that occurs is generally smaller than during the hard lockdown period. The lowest average value of change is equal to 52.64\% and the largest to 116.36\%. Despite the large change, values of VUF do not violate the 2\% limitation in any node and time period, with the largest value of 0.56\%.

As mentioned before, the value of THD$_u$ is largely affected by the situation at the fundamental frequency. Even though the results of the voltage magnitude analysis show that there are some nodes in which the average value of voltage magnitude increased, a more detailed analysis of changes in the post-lockdown period shows that there are more time periods with the decrease of voltage magnitude than with increase. Also, the value of THD$_u$ is dependent on more parameters than only on consumption and voltage magnitude at the fundamental frequency. The results in Figure \ref{fig:thd_pre_post} show that the average harmonic distortion increased in all nodes in the observed network, with all changes in the range between 1.51\% and 7.21\%, which are smaller values compared to the changes during the hard lockdown. Also, the maximum value of THD$_u$ is equal to 0.819\%, which is lower than the maximum that occurred during the hard lockdown but also during the pre-lockdown period. It is interesting to mention that the trend of the change of THD$_u$ in this scenario is almost identical to the trend of the voltage magnitude change. Nodes with the largest average increase of THD$_u$ are also the nodes with the average decrease of voltage magnitude and those with the lowest increase of THD$_u$ are those with the largest increase of voltage magnitude. The perceived trend additionally brings the value of the average harmonic distortion change into correlation with the average change of current and voltage magnitude at the fundamental frequency.

\section{Improvement of the PQ indicators}\label{sec:improvement}
\subsection{Methodology}
Even though changes in end-users' habits that arose as a consequence of restrictions during the pandemic have created numerous challenges for DSOs, they also enable potential for a larger exploitation of their flexibility in the planning and operation of distribution networks. In most cases, end-users can help a DSO by increasing or decreasing their consumption or by time-shifting of electricity consumption. Since there are papers that implement a demand-side management scheme and investigate if it is possible to improve the values of different PQ indicators only by increasing and decreasing end-users demand or test the time-shifting approach in a controlled laboratory environment, the focus in this paper is put on the comprehensive improvement of PQ in a real-world LV network only by shifting consumption to other time periods.

In order to avoid unnecessary time-shifting of consumption, happening too often, which will disrupt the comfort of end-users and discourage them from providing such and other similar flexibility services in the future, we developed an algorithm that detects time periods most suitable for time-shifting of consumption. The developed algorithm is a part of an integrated framework presented in this paper. For each day, an algorithm finds 10 time periods in which the change of electricity demand during the hard lockdown period was the largest. The change of electricity demand during hard and post-lockdown is defined with eq. (\ref{eq:detect}). Even though the trigger for activating the time-shifting service could be set in a different way, e.g., the change of demand was 30\% larger compared to pre-lockdown, such definition would lead to too often activation of the time-shifting of consumption service. Therefore, the limitation was set to 10 intervals with the highest value of change for each day.

\begin{equation}
    \Delta_{hard/post-pre} = \frac{P_{hard/post,t}-P_{pre,t}}{P_{pre,t}} 
    \label{eq:detect}
\end{equation}

After all such time periods are detected, the surplus of demand is shifted in eight consecutive time periods (2 hours) within 32 periods (8 hours) before or after the detected surplus. Time periods in which demand will be shifted are determined as those with the smallest total electricity demand since they will be most adequate for the increase of electricity demand. The step in which the minimum sum is found is defined with eq. (\ref{eq:min_P}).

\begin{equation}
\begin{gathered}
    \min \sum_{i = 0}^7 P_{t+i} \\ \forall t \in [T-32, T-8] \ or \ [T+1, T+24]
\end{gathered}
\label{eq:min_P}
\end{equation}

After the detection of the eight most suitable time periods $[t, t+7]$, the value of the demand increase in each time period $t+i$ is calculated using eq. (\ref{eq:delta_P}). Such an approach ensures fair distribution of the total increase of demand among consecutive time periods. An algorithm continuously updates the values of demand, i.e., if there are two consecutive periods detected as the most suitable for time-shifting of consumption, an algorithm will find adequate periods according to updated and not initial demand values.

\begin{equation}
\begin{gathered}
    \Delta_{t+i} = \frac{P_{t+j}}{\sum_{j = 0}^7 P_{t+i}} \cdot \frac{(P_{hard/post,t}-P_{pre,t})}{P_{pre,t}} 
    ,\quad \forall i \in [0,7]
\end{gathered}
\label{eq:delta_P}
\end{equation}

It is important to mention that only active power was observed in the described approach. Reactive power was calculated with the same power factor as in the initial case, before the improvement of the hard lockdown and post-lockdown consumption curves.

\subsection{Results}
After the implementation of the proposed methodology, the same analyses were made as in Section \ref{sec:results} with the modified, time-shifted consumption curve, which is compared with the results obtained using the pre-lockdown period. The results of that comparison will show if the time-shifting of consumption helped in the PQ improvement or if it did not have any effect and the values remained unchanged.

After the verification of an algorithm used for the detection of time intervals most adequate for time-shifting, more detailed analyses were conducted in order to investigate the effect of time-shifting of consumption on PQ indicators in LV distribution networks. The change of value of the PQ indicators is calculated with respect to the initial value, before the implementation of the demand time-shifting algorithm, using eq. (\ref{eq:change_imp}).

\begin{equation}
    \Delta_{indicator} = \frac{PQ_{indicator, improved} - PQ_{indicator, initial}}{PQ_{indicator, initial}}
    \label{eq:change_imp}
\end{equation}

Table \ref{tab:pq_hard_ts} shows the result of the change of voltage magnitude, VUF, and THD$_u$ after the implementation of the demand time-shifting algorithm for the hard lockdown period. The results present maximum increase, maximum decrease, average change, and median change for all defined PQ indicators. The maximum increase of voltage magnitude is around 4.5\% and the maximum value of voltage magnitude decrease is more than 3\%. Even though values of both median and average change are larger than zero, such change of the values that are in the interval between 90\% and 100\% are not so significant. The time-shifting of demand decreases peak demand and voltage magnitude in some intervals but consequentially increases both demand and voltage magnitude in other periods, causing a so called rebound effect. This effect does not present problems for DSOs since it cannot cause overvoltage or undervoltage, despite the increase in average and median voltage magnitude change. A more detailed analysis of results shows that there are not values close to the upper voltage limitation. 

Both VUF and THD$_u$ values are decreased in most periods, which can be seen from the negative median changes of the mentioned values. Same as in the case of voltage magnitude, due to the rebound effect, values of PQ indicators increase in some periods but the value of the increase is around ten times smaller than the increase in case of voltage magnitude. The maximum value of VUF is detected as an outlier value that does not adequately present the potential of the used methodology. Due to this and other similar outliers, the average value of the VUF change is larger than zero. Therefore, using a median value is a better indicatior of the improvement of PQ distrubances.

\begin{table}[htbp]
\caption{Change of PQ indicators during hard lockdown - time-shifted consumption}
\resizebox{\columnwidth}{!}{\begin{tabular}{c|c|c|c|c|}
\cline{2-5}
                                                                                           & \textbf{\begin{tabular}[c]{@{}c@{}}Maximum\\ increase (\%)\end{tabular}} & \textbf{\begin{tabular}[c]{@{}c@{}}Maximum\\ decrease (\%)\end{tabular}} & \textbf{\begin{tabular}[c]{@{}c@{}}Average\\ change (\%)\end{tabular}} & \textbf{\begin{tabular}[c]{@{}c@{}}Median\\ change (\%)\end{tabular}} \\ \hline
\multicolumn{1}{|c|}{\textbf{\begin{tabular}[c]{@{}c@{}}Voltage\\ magnitude\end{tabular}}} & 4.4458                                                                   & -3.1886                                                                  & 0.3715                                                                 & 0.2501                                                                \\ \hline
\multicolumn{1}{|c|}{\textbf{VUF}}                                                         & 26446.6319                                                                   & -99.156                                                                  & 16.9229                                                                & -9.5514                                                               \\ \hline
\multicolumn{1}{|c|}{\textbf{THD$_u$}}                                                         & 91.2088                                                                   & -48.3703                                                                  & -11.6702                                                                & -12.3757                                                               \\ \hline
\end{tabular}}
\label{tab:pq_hard_ts}
\end{table}

Since the measures related to the COVID-19 pandemic in most countries are seriously relaxed or even completely abandoned, the results related to the improvement of PQ indicators due to demand time-shifting in the hard lockdown period are no longer as relevant as they were a few months or years ago. However, some end-users' habits remain changed even after the hard lockdown and in this paper we test if these habits affect consumption in a way that the improvement of PQ after the implementation of the demand time-shifting algorithm is still possible.

As it can be seen in Table \ref{tab:pq_post_ts}, despite relaxing and abandoning  the COVID-19-related measures, end-users can still participate in the improvement of PQ indicators. Even though the values of change are smaller when compared to the potential of demand time-shifting in the hard lockdown period, trends of the voltage increase caused by the demand rebound effect, the median decrease VUF and THD$_u$ and outlier maximum value of VUF remain the same as in the hard lockdown period.

\begin{table}[htbp]
\caption{Change of PQ indicators during post-lockdown - time-shifted consumption}
\resizebox{\columnwidth}{!}{\begin{tabular}{c|c|c|c|c|}
\cline{2-5}
                                                                                           & \textbf{\begin{tabular}[c]{@{}c@{}}Maximum\\ increase (\%)\end{tabular}} & \textbf{\begin{tabular}[c]{@{}c@{}}Maximum\\ decrease (\%)\end{tabular}} & \textbf{\begin{tabular}[c]{@{}c@{}}Average\\ change (\%)\end{tabular}} & \textbf{\begin{tabular}[c]{@{}c@{}}Median\\ change (\%)\end{tabular}} \\ \hline
\multicolumn{1}{|c|}{\textbf{\begin{tabular}[c]{@{}c@{}}Voltage\\ magnitude\end{tabular}}} & 2.8399                                                                   & -2.4235                                                                  & 0.2728                                                                  & 0.1811                                                                \\ \hline
\multicolumn{1}{|c|}{\textbf{VUF}}                                                         & 17230.1585                                                                   & -99.8613                                                                  & 7.7298                                                                & -8.8347                                                               \\ \hline
\multicolumn{1}{|c|}{\textbf{THD$_u$}}                                                         & 76.0736                                                                   & -46.7923                                                                  & -9.9796                                                                & -10.1633                                                               \\ \hline
\end{tabular}}
\label{tab:pq_post_ts}
\end{table}

Improvement of PQ by implementing demand-shifting algorithm is achieved both in the hard lockdown period but also in more recent post-lockdown period. Although the improvement is less significant for the consumption curve in the post-lockdown period, values of the observed changes show that the implementation of the presented algorithm was not useful only during hard lockdown but even in the case of relaxed and abandoned measures. 

There are many other proposed approaches for the improvement of values of PQ indicators, and many of them include the implementation of physical devices such as voltage regulators, phase switching devices, or active and passive harmonic filters. Even though their efficiency is high and they provide the possibility to improve certain technical conditions without compromising the comfort of end-users, their integration in a network often requires investment creating additional and potentially unwanted costs for end-users. Based on the results and the efficiency of the tested approach, the demand time-shifting solution proposed in this paper can be considered as an alternative method for mitigating disturbances in LV distribution networks. However, the proposed method also requires the replacement of the metering equipment, the development of control algorithms, and the installation of equipment needed for receiving and answering signals for decreasing or increasing electricity demand.

\section{Conclusion}\label{sec:conclusion}
Together with the rapid integration of LC units, the changes in end-users' habits, such as those exhibited during the pandemic lockdowns, present the biggest challenge in the planning and operation of end-users that the DSOs faced recently. To overcome these challenges, now more than ever, there is an emerging need for the development of tools that can be used both in the short-term and long-term planning but also in the everyday operation of distribution networks affected by the penetration of LC units and long-term changes caused by the pandemic.

One of such tools is presented in this paper, where we rely on functionalities of open source technologies Python, SQL with the postgis extension, and QGIS, and develop a GIS-based framework that is used for processing GIS and smart meters data, creation of an LV network's mathematical model, power quality analyses, and finally, the visualisation of the simulation results. In the first step, several typical errors in GIS and smart meters data are identified and removed using Python and SQL with postgis extension so they can be used in the second step, creating a mathematical model of a network. A mathematical model of a network is created with GIS data containing both technical and geospatial attributes and representing the elements of distribution networks such as substations, lines, or electric switch cabinets. A mathematical model is created with pandapower, the Python-based simulation tool. Pandapower is also used in the third step, where it is used together with the harmonic analysis extension, developed by the authors of this paper, for power quality simulations. PQ indicators voltage magnitude, VUF, and THD$_u$ are calculated and the average change of their value during the hard and post-lockdown periods are compared to the values in the initial, pre-lockdown periods. The final step of the developed tool includes a GIS visualisation with the QIGS open source tool. The results show that on average, all PQ indicators deteriorated during both pandemic periods, with larger disturbances in the hard lockdown period. 

According to calculated values, it is possible to conclude that increased consumption during the pandemic presents a potential problem for the DSOs, especially in the networks that are expected to integrate new LC units. To overcome the potential issue, an algorithm that detects time periods most suitable for the shifting of end-users' consumption is implemented as an extension of the presented framework. An algorithm creates a new consumption curve, with minimally discomforting end-users. An implementation of time-shifting of consumption successfully decreased PQ disturbances, with even smaller deterioration of PQ indicators than in the pre-lockdown period. 

The results and conclusions presented in the paper are valid for only one analysed LV feeder and they are not relevant to the analysis of all LV distribution networks in the area. However, the main contribution of the developed GIS-based open source tool is the possibility given to a user who can analyse a variety of different feeders or a whole network in order to draw more general conclusions. The feeder in this paper is used as a proof of concept and a test case used for verification of the developed tool. Moreover, the conclusions related to the potential of end-users’ flexibility in the PQ improvement are valid no matter the type of network, the only thing that changes is the extent to which end-users can help in mitigating PQ deterioration.

Future work includes further investigation of the potential of using both end-users' flexibility and utilization of physical devices, how they affect each other's operation, and how they affect the operation costs of DSOs caused by PQ disturbances but also on the finances of end-users considering the incentives for providing different services. The emphasis of future work is planned to be on investigating the economic feasibility of integrating different solutions for the improvement of PQ indicators, including a detailed economic analysis of the ways through which end-users can be financially encouraged to provide services that will lead to a safer and more reliable operation of LV networks.

\bibliographystyle{IEEEtran} 
\bibliography{cas-refs}

\begin{thebibliography}{10}
\providecommand{\url}[1]{#1}
\csname url@samestyle\endcsname
\providecommand{\newblock}{\relax}
\providecommand{\bibinfo}[2]{#2}
\providecommand{\BIBentrySTDinterwordspacing}{\spaceskip=0pt\relax}
\providecommand{\BIBentryALTinterwordstretchfactor}{4}
\providecommand{\BIBentryALTinterwordspacing}{\spaceskip=\fontdimen2\font plus
\BIBentryALTinterwordstretchfactor\fontdimen3\font minus \fontdimen4\font\relax}
\providecommand{\BIBforeignlanguage}[2]{{%
\expandafter\ifx\csname l@#1\endcsname\relax
\typeout{** WARNING: IEEEtran.bst: No hyphenation pattern has been}%
\typeout{** loaded for the language `#1'. Using the pattern for}%
\typeout{** the default language instead.}%
\else
\language=\csname l@#1\endcsname
\fi
#2}}
\providecommand{\BIBdecl}{\relax}
\BIBdecl

\bibitem{TORRES2022117718}
S.~Torres, I.~Durán, A.~Marulanda, A.~Pavas, and J.~Quirós-Tortós, ``{Electric vehicles and power quality in low voltage networks: Real data analysis and modeling},'' \emph{Applied Energy}, vol. 305, p. 117718, 2022.

\bibitem{WANG2019113927}
L.~Wang, R.~Yan, and T.~K. Saha, ``{Voltage regulation challenges with unbalanced PV integration in low voltage distribution systems and the corresponding solution},'' \emph{Applied Energy}, vol. 256, p. 113927, 2019.

\bibitem{WANG2020115735}
Q.~Wang, M.~Lu, Z.~Bai, and K.~Wang, ``{Coronavirus pandemic reduced China’s CO2 emissions in short-term, while stimulus packages may lead to emissions growth in medium- and long-term},'' \emph{Applied Energy}, vol. 278, p. 115735, 2020.

\bibitem{RUAN2021116354}
G.~Ruan, J.~Wu, H.~Zhong, Q.~Xia, and L.~Xie, ``{Quantitative assessment of U.S. bulk power systems and market operations during the COVID-19 pandemic},'' \emph{Applied Energy}, vol. 286, p. 116354, 2021.

\bibitem{COSTA2021107172}
V.~B. Costa, B.~D. Bonatto, L.~C. Pereira, and P.~F. Silva, ``{Analysis of the impact of COVID-19 pandemic on the Brazilian distribution electricity market based on a socioeconomic regulatory model},'' \emph{International Journal of Electrical Power \& Energy Systems}, vol. 132, p. 107172, 2021.

\bibitem{MAHFUZALAM2021106832}
S.~{Mahfuz Alam} and M.~H. Ali, ``{Analysis of COVID-19 effect on residential loads and distribution transformers},'' \emph{International Journal of Electrical Power \& Energy Systems}, vol. 129, p. 106832, 2021.

\bibitem{ROULEAU2021116565}
J.~Rouleau and L.~Gosselin, ``{Impacts of the COVID-19 lockdown on energy consumption in a Canadian social housing building},'' \emph{Applied Energy}, vol. 287, p. 116565, 2021.

\bibitem{8344496}
L.~Thurner, A.~Scheidler, F.~Schäfer, J.-H. Menke, J.~Dollichon, F.~Meier, S.~Meinecke, and M.~Braun, ``{Pandapower—An Open-Source Python Tool for Convenient Modeling, Analysis, and Optimization of Electric Power Systems},'' \emph{IEEE Transactions on Power Systems}, vol.~33, no.~6, pp. 6510--6521, 2018.

\bibitem{ANTIC2022107935}
T.~Antić, L.~Thurner, T.~Capuder, and I.~Pavić, ``Modeling and open source implementation of balanced and unbalanced harmonic analysis in radial distribution networks,'' \emph{Electric Power Systems Research}, vol. 209, p. 107935, 2022.

\bibitem{MORSTYN2020115397}
T.~Morstyn, K.~A. Collett, A.~Vijay, M.~Deakin, S.~Wheeler, S.~M. Bhagavathy, F.~Fele, and M.~D. McCulloch, ``{OPEN: An open-source platform for developing smart local energy system applications},'' \emph{Applied Energy}, vol. 275, p. 115397, 2020.

\bibitem{5491276}
R.~D. Zimmerman, C.~E. Murillo-Sánchez, and R.~J. Thomas, ``Matpower: Steady-state operations, planning, and analysis tools for power systems research and education,'' \emph{IEEE Transactions on Power Systems}, vol.~26, no.~1, pp. 12--19, 2011.

\bibitem{GisOpenDSS}
G.~Valverde, A.~Arguello, R.~González, and J.~Quirós-Tortós, ``Integration of open source tools for studying large-scale distribution networks,'' \emph{IET Generation, Transmission \& Distribution}, vol.~11, no.~12, pp. 3106--3114, 2017.

\bibitem{ZAMBRANOASANZA2021110853}
S.~Zambrano-Asanza, J.~Quiros-Tortos, and J.~F. Franco, ``{Optimal site selection for photovoltaic power plants using a GIS-based multi-criteria decision making and spatial overlay with electric load},'' \emph{Renewable and Sustainable Energy Reviews}, vol. 143, p. 110853, 2021.

\bibitem{8270364}
R.~Torquato, D.~Salles, C.~Oriente~Pereira, P.~C.~M. Meira, and W.~Freitas, ``{A Comprehensive Assessment of PV Hosting Capacity on Low-Voltage Distribution Systems},'' \emph{IEEE Transactions on Power Delivery}, vol.~33, no.~2, pp. 1002--1012, 2018.

\bibitem{HUANG2019113855}
P.~Huang, Z.~Ma, L.~Xiao, and Y.~Sun, ``{Geographic Information System-assisted optimal design of renewable powered electric vehicle charging stations in high-density cities},'' \emph{Applied Energy}, vol. 255, p. 113855, 2019.

\bibitem{9248930}
E.~Vega-Fuentes, J.~Yang, and C.~Lou, ``{Power Flow Visualization in DER-Rich Low Voltage Networks},'' in \emph{2020 IEEE PES Innovative Smart Grid Technologies Europe (ISGT-Europe)}, 2020, pp. 735--738.

\bibitem{8493353}
A.~Guzmán, A.~Argüello, J.~Quirós-Tortós, and G.~Valverde, ``{Processing and Correction of Secondary System Models in Geographic Information Systems},'' \emph{IEEE Transactions on Industrial Informatics}, vol.~15, no.~6, pp. 3482--3491, 2019.

\bibitem{KU2022118539}
A.~L. Ku, Y.~L. Qiu, J.~Lou, D.~Nock, and B.~Xing, ``{Changes in hourly electricity consumption under COVID mandates: A glance to future hourly residential power consumption pattern with remote work in Arizona},'' \emph{Applied Energy}, vol. 310, p. 118539, 2022.

\bibitem{QI2020115708}
N.~Qi, L.~Cheng, H.~Xu, K.~Wu, X.~Li, Y.~Wang, and R.~Liu, ``Smart meter data-driven evaluation of operational demand response potential of residential air conditioning loads,'' \emph{Applied Energy}, vol. 279, p. 115708, 2020.

\bibitem{FEKRI2021116177}
M.~N. Fekri, H.~Patel, K.~Grolinger, and V.~Sharma, ``{Deep learning for load forecasting with smart meter data: Online Adaptive Recurrent Neural Network},'' \emph{Applied Energy}, vol. 282, p. 116177, 2021.

\bibitem{YILDIZ2017402}
B.~Yildiz, J.~Bilbao, J.~Dore, and A.~Sproul, ``Recent advances in the analysis of residential electricity consumption and applications of smart meter data,'' \emph{Applied Energy}, vol. 208, pp. 402--427, 2017.

\bibitem{8322199}
Y.~Wang, Q.~Chen, T.~Hong, and C.~Kang, ``Review of smart meter data analytics: Applications, methodologies, and challenges,'' \emph{IEEE Transactions on Smart Grid}, vol.~10, no.~3, pp. 3125--3148, 2019.

\bibitem{6408380}
A.~Kulkarni and V.~John, ``{Mitigation of Lower Order Harmonics in a Grid-Connected Single-Phase PV Inverter},'' \emph{IEEE Transactions on Power Electronics}, vol.~28, no.~11, pp. 5024--5037, 2013.

\bibitem{8637800}
N.~Kumar, B.~Singh, B.~K. Panigrahi, C.~Chakraborty, H.~M. Suryawanshi, and V.~Verma, ``{Integration of Solar PV With Low-Voltage Weak Grid System: Using Normalized Laplacian Kernel Adaptive Kalman Filter and Learning Based InC Algorithm},'' \emph{IEEE Transactions on Power Electronics}, vol.~34, no.~11, pp. 10\,746--10\,758, 2019.

\bibitem{CHAMINDABANDARA2020116022}
W.~{Chaminda Bandara}, G.~Godaliyadda, M.~Ekanayake, and J.~Ekanayake, ``{Coordinated photovoltaic re-phasing: A novel method to maximize renewable energy integration in low voltage networks by mitigating network unbalances},'' \emph{Applied Energy}, vol. 280, p. 116022, 2020.

\bibitem{THOMAS2020114314}
D.~Thomas, G.~D’Hoop, O.~Deblecker, K.~N. Genikomsakis, and C.~S. Ioakimidis, ``An integrated tool for optimal energy scheduling and power quality improvement of a microgrid under multiple demand response schemes,'' \emph{Applied Energy}, vol. 260, p. 114314, 2020.

\bibitem{CICEK2021106528}
A.~Çiçek, A.~K. Erenoğlu, O.~Erdinç, A.~Bozkurt, A.~Taşcıkaraoğlu, and J.~P. Catalão, ``Implementing a demand side management strategy for harmonics mitigation in a smart home using real measurements of household appliances,'' \emph{International Journal of Electrical Power \& Energy Systems}, vol. 125, p. 106528, 2021.

\bibitem{VIJAYAN2021117361}
V.~Vijayan, A.~Mohapatra, and S.~Singh, ``Demand response with volt/var optimization for unbalanced active distribution systems,'' \emph{Applied Energy}, vol. 300, p. 117361, 2021.

\bibitem{PAPAIOANNOU2013540}
I.~T. Papaioannou, A.~Purvins, and E.~Tzimas, ``Demand shifting analysis at high penetration of distributed generation in low voltage grids,'' \emph{International Journal of Electrical Power \& Energy Systems}, vol.~44, no.~1, pp. 540--546, 2013.

\end{thebibliography}

\end{document}